\documentclass[fleqn,usenatbib,useAMS]{mnras}

\usepackage[utf8]{inputenc}
\usepackage{graphicx}	
\usepackage{amsmath}	
\usepackage{amssymb}	
\usepackage{multicol}        
\usepackage{bm}		
\usepackage{pdflscape}	

\newcommand{\Msol}{\mathrm{M}_{\odot}} 
\newcommand{\Mpc}{\mathrm{Mpc}} 
\newcommand{\de}{\mathrm{d}} 
\newcommand{\sigmacrit}{\Sigma_{crit}} 

\usepackage[T1]{fontenc}
\usepackage{ae,aecompl}

\usepackage{newtxtext,newtxmath}

\usepackage{hyperref}

\title[Concentrations \& Magnitude Gaps in CS82]{On Mass Concentrations \& Magnitude Gaps of Galaxy Systems in the CS82 Survey}

\author[Vitorelli et al.]{André Z. Vitorelli$^{1}$\thanks{Contact e-mail: \href{mailto:andre.vitorelli@usp.br}{andre.vitorelli@usp.br}}, 
    Eduardo S. Cypriano$^{1}$, 
    Martín Makler$^{2}$,
    Maria E. S. Pereira$^{2}$,\newauthor
    Thomas Erben$^{3}$, 
    Bruno Moraes$^{4}$.
\\
$^{1}$Instituto de Astronomia Geofísica e Ciências Atmosférias, Universidade de São Paulo, Cidade Universitária, 05508-090, São Paulo, Brazil \\
$^{2}$Centro  Brasileiro  de  Pesquisas  Físicas,  Rua  Dr. Xavier Sigaud 150, CEP 22290-180, Rio de Janeiro, RJ, Brazil \\
$^{3}$Argelander-Institut fur Astronomie, Auf dem Hugel 71, D-53121 Bonn, Germany\\
$^{4}$Dept. of Physics and Astronomy, University College London, London, WC1E 6BT \\ \\
}

\date{Last updated: \today}

\pubyear{2017}

\begin{document}

\label{firstpage}
\pagerange{\pageref{firstpage}--\pageref{lastpage}}
\maketitle

\begin{abstract}
Galaxy systems with large magnitude gaps - defined as the difference in magnitude between the central galaxy and the brightest satellite in the central region - have been claimed to have earlier formation histories. In this study we measure the mass concentration, as an indicator of early formation, of ensembles of galaxy systems divided by redshift and magnitude gaps in the $r$ band. We use cross-correlation weak lensing measurements with NFW parametric mass profiles to measure masses and concentrations of these ensembles from a catalogue of systems built from the SDSS Coadd by the redMaPPer algorithm. The lensing shear data come from the CFHT Stripe 82 (CS82) survey, and consists of $i$-band images of the SDSS Stripe 82 region. We find that the stack made up of systems with larger magnitude gaps  has a high probability of being more concentrated on average, in the lowest redshift slice ($0.2<z<0.4$), both when dividing in quartiles ($P=0.98$) and terciles ($P=0.85$). These results lend credibility to the claim that the magnitude gap is an indicator of earlier formed systems. 
\end{abstract}

\section{Introduction}

In gravitationally bound systems of galaxies, both the larger clusters or smaller groups, the study of the relationship between the central galaxy (CG) and its host system properties is important to understand the evolution of these structures and their value as cosmological probes \citep{DubinskiBCG98,AnjaBCG2007,Kravtsov2012}. The predominance of the CG over the system can be represented in several ways, but is most commonly measured by the difference in magnitude of the CG, which is usually the most luminous, to the second brightest in a central region (BSG - Brightest Satellite Galaxy).

This \emph{magnitude gap} has been suggested as an indicator of early formation of the system, in the sense of accreting more than half of its mass at high redshifts. This is because it is thought that these large magnitude gaps develop through dynamical processes that result in the CG absorbing massive satellite galaxies \citep{Jones2000,Jones03}, while the lack of subsequent mergers leave the system depleted of bright galaxies except for the CG.

The first identification of a system displaying a large discrepancy between the magnitude of the central galaxy and its companions was made by \citet{Ponman94}, when they investigated the system $\mathrm{RX \ J}1340.6+4018$. At first, it was thought to consist of a single galaxy bearing an overextended X-ray halo, tracing a deep gravitational potential well. It was later shown by \citet{Jones2000}, however, to be actually a group of about $N\sim 10$ galaxies, out of which the central galaxy accounts for about $\sim 70\%$ of the total optical luminosity. Under the light of the then nascent field of galaxy formation simulations \citep{Barnes89,Bode93}, the aforementioned early formation hypothesis was proposed.

An empirical definition was given afterwards by \citep{Jones03} for what was then named \emph{fossil groups} (FGs), and is expressed by

\begin{enumerate}
 \item a high luminosity in X-rays $L_X \geq 0.25 \times 10^{42} \mathrm{erg} \ \mathrm{s}^{-1}$,
 \item an absolute magnitude gap between the central, usually most luminous galaxy and the second brightest galaxy greater than $\Delta M_{1-2}= M_{BSG}-M_{BCG} \geq 2$ within half the projected $r_{200}$ radius. 
\end{enumerate}

These criteria single out extreme cases of systems where satellite galaxies are orders of magnitude less luminous than the CG. The extended diffuse X-Ray halo, in turn, indicates a potential well which would be expected to be filled with brighter galaxies.

Further observational studies offered additional insights on the characteristics of FGs. \citet{ClaudiaCyprianoLaerte05,Cypriano2006} have shown that large magnitude gap systems exist also in higher mass systems, proposing them as \emph{Fossil Clusters}\footnote{Putting together FGs and FCs, we label them as \emph{Fossil Systems} (FSs) in this work.}. \citet{Khosroshahi2004} studied the shape of the CGs of fossil groups, showing indicatives of wet mergers, while \citet{Khosroshahi2007} studied a few systems indicating that their scaling  relations did not significantly differ from groups of the same mass. The more recent FOGO (Fossil Group Origins - \citet{FOGO1,FOGO2,FOGO3,FOGO4,FOGO5,FOGO6,FOGO7}) project has studied fossil groups in many aspects, from a single system to larger ($n=102$) samples, having found no differences between the scaling relations of fossil groups and that of non-fossils of comparable mass, being consistent with the CG accreting of satellite galaxies scenario. However, analysing substructures, they found that FGs are not as relaxed as expected from simulations.

Using N-body simulations, \citet{Dariush07} claimed that fossil groups accrete most of their mass earlier on average, lending support to the early formation history hypothesis, while \citet{vonBendaBeckmann2008} have shown that fossil groups accreted on average only one further bright galaxy since $z=1$, compared to the average of three for other groups. In hydrodynamic simulations, \citet{Donghia2005} found that FGs assembled half of their masses before $z=1$.

The idea that large magnitude gaps indicate early halo formation is not undisputed, however. At the very least, it can be shown that random draws from Schechter function for groups of fewer galaxies have a higher probability of yielding larger magnitude gaps in the bright end (e.g. see \citet{Hearin2013}, Sec. 4). Furthermore, \citet{ParanjapeSheth12} have shown that typical luminosities for both CGs and BSG in a SDSS catalogue of clusters are compatible with draws from the brightest and second brightest galaxy in the same luminosity function.\citet{MulchaeyZabludoff}, analysing NGC 1132, have suggested that fossil groups may consist of \emph{failed groups}, that is, local overdensities in which other bright galaxies never formed. The masses and $M/L$ ratios of the central galaxies of FSs are also usually too large to be explained as end points of compact group evolution driven just by dynamical friction \citep{Voevodkin2010}. Finally \citet{Proctor11,Harrison2012,FOGO3} find evidence in support of this scenario, indicating that FGs might just be the result of massive halos with low occupation number.

Another consideration is that groups and clusters may go into a fossil phase in their history, with an absence of significant mergers \citep{vonBendaBeckmann2008}. A bright field galaxy, or even a nearby group, may then be accreted and then replenish the inner region with bright galaxies again. In simulations, \citet{Dariush10} find that about $90\%$ of fossil groups that were identified in earlier epochs become non fossils after $4\mathrm{Gyr}$ and that the fossil phase persists for $\sim 1\mathrm{Gyr}$. Using semi-analytic models based on the Millennium simulation \citep{Millenium}, and \citet{Gozaliasl14} have shown that $80\%$ of groups ($13<\log M_{200} < 14$ in $\Msol$) that would be classified as fossil at redshift $z=1$, thereby lose their large magnitude gaps with time, but that $40\%$ of the clusters ($\log M_{200} > 14$ in $\Msol$), on the other hand, retain large gaps.

With the given definition, FGs are not uncommon in the local universe, with estimates of their population in the local universe varying from $\sim 5\%$ to $40\%$ of all \citep{vonBendaBeckmann2008}, depending on selection. \citet{Donghia2005} finds about $33 \%$ FGs whereas the common literature figure hovers around $10\sim20\%$ of the overall group population. Statistical comparisons between populations of observed fossil systems and non-fossil galaxy associations are still lacking due to the need of larger number of fossil systems and the fact that their lower percentage on the existing X-ray surveys do not provide large enough samples. Recent simulation works show that the magnitude gap by itself is not enough to produce a pure set of early forming systems \citep{Deason2013}, and suggested that contamination by other factors can be removed by studying the relationship between the CG position and the centroid of the luminosity distribution of the system as a whole \citep{Raouf16}.

In order to better understand the relationship between the magnitude gap and formation history, we investigate its correlation with mass concentrations, using cross-correlation (also called \emph{stacking} of) weak-lensing methods with parametric mass profiles. Previous analysis both in observations \citep{Khosroshahi2007} and simulations \citep{Deason2013} indicate that larger magnitude gap will correspond, on average, to more concentrated systems, as is expected for the lower-mass earlier formed systems \citep{NFW97}. However, previous observational studies measured concentrations on a handful ($n\sim10$) of systems at a time. We want to go further by using a survey with a large number of systems to study the average concentration in populations ranked by $\Delta M_{1-2}$.

Instead of focusing on optically selected Fossil Systems specifically as a separated population, we define ranked \emph{stacks} by magnitude gaps, after dividing our whole population in redshift \emph{slices}, and measure their mass concentrations through stacking their weak lensing signal, using a parametric profile. 

In section 2 we present our data sources, followed by a description of the  cross-correlation weak-lensing method. In section 4 we describe the MCMC fitting method used and present the results for the masses and concentrations of each stack. Finally, in section 5 we analyse the difference in concentration found between stacks followed by a discussion of our results.

Throughout this paper we use the standard Flat $\Lambda$CDM cosmology with $h_{100}=0.7$, $\Omega_m=0.3$. 

\section{Data}

The data used in this work come from two different surveys: a catalogue of galaxy clusters built by the \citet{redMaPPer} algorithm on SDSS images \citep{Coadd} and the CFHT Stripe-82 survey (CS82; \citet{cs82}). The so called Stripe-82 is an equatorial region about $2^{\circ}$ wide in latitudes between $-40^{\circ}<\mathrm{RA} <60^{\circ}$, which has been extensively investigated in many different bands from various instruments \citep{HerS2014,S82X,Clusters82}. The CS82 was then specifically designed to take profit of the synergies in this abundance of data, adding shear measurements for weak lensing analysis of the large scale structure of the universe. 

\subsection{The redMaPPer catalogue}
The catalogue of galaxy clusters and groups used was obtained by the redMaPPer (\textbf{red}-sequence \textbf{M}atched-filter \textbf{P}robabilistic \textbf{Per}colation) algorithm for cluster identification. redMaPPer is a photometric cluster finding algorithm based on the optimised richness estimator $\lambda$ of \citet{RichnessEstimator2012}, which is designed to have minimum scatter with cluster masses.

The redMaPPer algorithm identifies galaxy clusters as overdensities of \emph{red-sequence galaxies} around central galaxy candidates. First it uses learning techniques on spectroscopic training sets to characterise the evolution of the red sequence as a function of redshift. It then uses the resulting red sequence model, together with a radial aperture filter and a luminosity function filter based on the Schechter function to estimate the probability that any given observed galaxy belongs to some cluster. The cluster richness is then defined as the sum of the probabilities of galaxies considered, as an estimator for the expected value of the number of galaxies in the cluster,

\begin{equation}
 \lambda = \sum_i p_i\,.
\end{equation}

In addition, by identifying a red-sequence for each cluster, it can estimate a cluster redshift by simultaneous fitting all possible member galaxies to its red-sequence model. The aperture filter also defines a percolation radius $R_c$ that is related to the obtained richness $\lambda$ by

\begin{equation}
 R_c = 1 ~\Mpc \times \left( \frac{\lambda}{100} \right)^\beta\,,
\end{equation} with $\beta = 0.2$ \citep{redMaPPer,RichnessEstimator2012}.

 We use this radius to define the central region of the system as $R < R_c/3$, which translates into about $\sim 1/2 \ R_{200}$ \citep{RichnessEstimator2012}. We have attempted more restrictive radii considerations, to compensate for the large errors in magnitude gaps (due to the probabilistic nature of the cluster/group identification) but we found no qualitative differences in the results. 

\subsection{CS82 Data}
The CS82 survey consists of 173 pointings of the MegaCam instrument, using the $i.MP9702$ filter ($\sim$ SDSS $i$ band). The MegaCam is a large ($1 \mathrm{deg}^2$) field of view camera with an angular scale of $0.187 \mathrm{arcsec}/\mathrm{pixel}$.  The completeness magnitude limit achieved was $m_i < 24$, with an excellent median seeing of $\sim 0.6^{\prime \prime}$. The image reduction process profits from the CFHTLS pipeline. The total effective area after masking and de-overlapping the images corresponds to about $124 \deg^2$ of the sky. This magnitude limit is defined as a safe limit to guarantee homogeneity for all the 173 tiles, with an average galaxy density per image area of $\sim 10 \mathrm{gal} / \mathrm{arcmin}^2$.

The classification and measurements of shapes of objects has been done by the LensFit algorithm \citep{Lensfit1,Lensfit2,Lensfit3}, and the details of the calibration and its systematics are discussed in \citet{CFHTLenS2013}. In this work, all objects with magnitudes $i_{AB}<23.5$, with LensFit weight\footnote{The LensFit weight is a measure of the uncertainty in measured shapes of individual objects.} $w>0$ and $\mathrm{FITCLASS}$\footnote{$\mathrm{FITCLASS}$ is a star/galaxy identification parameter where $0$ corresponds to galaxies and $1$ corresponds to stars.} $=0$ are used. Together, this criteria result in a total of $4,450,478$ galaxies used for weak-lensing. The lensing products of this survey have been used previously in published results by \citet{HuanYuan14,Liu2015,HuanYuan15,Leauthaud17,Niemiec2017,RanLi2014,RanLi2016,Hand2015,Maria2017,Niemiec2017}.

Individual photometric redshifts for galaxies that were used in the shear analysis were taken from \citet{Bundy15}, and were calculated from SDSS photometric data by applying the BPZ algorithm of \citet{Benitez2000}.

\section{Method}

\subsection{Determination of Magnitude Gaps}
The calculation of magnitude gaps require that we address two important particularities of the redMaPPer catalogue: first the algorithm does not always identify the central galaxy as the brightest, which is in agreement with the literature \citep{Hoshino2015,Skibba2011}. Therefore, we refer to the central galaxy as CG, instead to the literature usual BCG, and the magnitude gaps are calculated between the most probable central galaxy and the brightest satellite galaxy  ($\Delta M_{1-2} := M_{BSG} - M_{CG}$ ). Second, redMaPPer calculates \emph{probabilities of memberships} for galaxies - that is, we cannot calculate $\Delta M_{1-2}$ straightforwardly by subtracting magnitudes of the CG and the the possible BSG in the catalogue, since we do not know with absolute certainty which galaxies are system members. Instead, we calculate \emph{expected values} of magnitude gaps, by simply computing the the expected value from the \emph{probability distribution} of the probable members catalogue of each system.

To find the expected value for the magnitude gap, we consider a list of all galaxies inside an inner region (as previously mentioned) of the system except for the central, with $N$ redMaPPer-identified possible members, so that their membership probabilities are given by  $\vec{p} = p_1, p_2, p_3 ...$ ordered by decreasing brightness (increasing absolute magnitude). Then, the probability of the first galaxy to be the brightest in this group is just $p_1$ as no other galaxy can possibly be brighter than it, So that $p_1=P_1$ is also the probability of the magnitude gap $\Delta M_{1-2}$ to be $M_{CG}-M_1$. Now, for the second galaxy we need to ensure that the first is not present $(1-p_1)$ \emph{and} that the second is, that is, $(1-p_1)p_2$. By simple iteration, the probability of the $n$-th galaxy of being the BSG is given by:

\begin{equation}
 P_n = p_n \prod_{i=2}^{n-1} (1-p_i)\,.
\end{equation}

Collecting these results the expected values are the sum weighted by individual probabilities for each possible gap:
\begin{equation}
E\left[\Delta M_{1,2} \right] = M_{CG}-\sum_{n=1}^N P_n M_n\,,  
\end{equation} and the errors adopted follow directly from the definition of the variance also,

\begin{equation}
 \mathrm{Var}\left[\Delta M_{1,2}\right] = E\left[\left(\Delta M_{1,2}\right)^2 \right] - \left[E\left(\Delta M_{1,2} \right)\right]^2\,.
\end{equation}

With that settled, we turn our attention to the mass distribution model.

\subsection{Parametric Modelling of Mass Profiles}

We now turn to details of the mass profile used, reviewing previous work from several groups \citep{Mandelbaum2008, Johnston2007,FordShear,HuanYuan15} to make any model choices explicit. Parametric modelling has the advantage of breaking the mass-sheet degeneracy through postulating spherical symmetry, which, even if not true in many galaxy systems, will be a good description for stacked data measured from randomly oriented systems.

The main component of the mass distribution of a galaxy system is the radial profile of the dark-matter dominated halo, which has been shown to follow roughly a universal form \citep{HussJainSteinmetz,Katz1991,ColeLacey96,NFW97}. Among several proposed expressions for the radial mass density, we use the simulation derived result by \citet{NFW97}, given by 

\begin{equation}
 \rho_{\mathrm{NFW}}(r)= \frac{\delta_{\mathrm{NFW}} \ \rho_{\mathrm{crit}}(z)}{(r/r_\mathrm{s})(1+r/r_\mathrm{s})^2}\,,
\end{equation} with

\begin{equation}
 \delta_{\mathrm{NFW}} = \frac{\Delta_{c}}{3}\frac{c_{\mathrm{N}}^3}{\ln(1+c_{\mathrm{N}})-c_{\mathrm{N}}/(1+c_{\mathrm{N}})}\,.
\end{equation} where $c$ is the concentration parameter, and $\Delta_c=200$ is the overdensity at collapse. 

Since the mass of the NFW profile diverges when the profile is integrated to infinity, we choose a cut-off radius at $R_{200}$, defined as the radius in which the average matter density is $200$ times the average density of matter in the universe. With that, the mass is given by

\begin{equation}
M_{200} = \Delta_c \Omega_M \bar{\rho}_c \frac{4}{3}\pi R_{200}^3    
\end{equation} which, together with the concentration parameter, labelled accordingly as $c_{200}$, has analytic expressions for the lensing shear \citep{OxacaBrainerd}. 

We can then refine our model by taking into account the probability of miscentering of a fraction of the systems included in a stack, which damp the signal in the inner regions and the additional mass resulting from the clustering of clusters at large radii.

Using the lensing differential surface mass density $\Delta \Sigma(R) = \bar{\Sigma}(r<R)-\Sigma(R)$, where $\bar{\Sigma}(R)$ is the average mass density inside a radius $R$ , we can parametrise the total lensing signal as

\begin{multline}\label{eq:fullmodel}
 \Delta \Sigma(R|M_0,M_{200},c_{200},p_{\mathrm{cc}},\sigma_{\mathrm{off}})= \frac{M_0}{\pi R^2} + p_{\mathrm{cc}}\Delta \Sigma_{\mathrm{NFW}}(R) + \\ (1-p_{\mathrm{cc}})\Delta \Sigma_{\mathrm{NFW}}^{\mathrm{off}}(R) + \Delta \Sigma_{2ht}(R)\,.
\end{multline}

Here the first term on the right-hand side of the equation was proposed by \citet{MandelbaumPHD} to account for the baryonic mass ($M_0$) of the central galaxy, the second accounts for the fraction ($p_{\mathrm{cc}}$) of correctly centred systems, the third for the miscentered fraction and the last one is the contribution of the large scale structure \citep{Johnston2007}. We now briefly describe the contributions of miscentering and  large scale structure to this model.

\subsubsection{Treating miscentered systems}

When using radially symmetric profiles to model a stack of mass distributions, it is important to consider how the overall profile is affected when a fraction of the observed systems may be incorrectly centred. It is known that some fraction of the CGs may be offset from the true centre of the gravitational potential it inhabits \citep{Girardi97,kkk1999}, and also that cluster finder algorithms may identify a wrong galaxy as the CG \citep{Johnston2007}. When we combine several clusters, the effect of this \emph{miscentering} is to produce lower levels of shear in the inner radii, as the density peak of the miscentered systems will be spread away from the centre, which may bias results towards lower halo masses \citep{Johnston2007}. 

If the $2D$ offset in the lens plane of a single profile is given by $R_{\mathrm{off}}$, the azimuthally averaged surface profile will be given by a shift of the centre and an integral around the correct centre as \citep{YangMiscentering2006}:

\begin{equation}
\Sigma_{\mathrm{off}} (R | R_{\mathrm{off}}) = \frac{1}{2 \pi} \int_0^{2 \pi} \de \theta  \ \Sigma \left(\sqrt{R^2 + R_{\mathrm{off}}^2 - 2 R R_{\mathrm{off}} cos(\theta)}\right)\,.
\end{equation} with the distribution of offsets being modelled as 

\begin{equation}
 P\left(R_{\mathrm{off}}\right) = \frac{R_{\mathrm{off}}}{\sigma_{\mathrm{off}}^2} \exp\left [-\frac{1}{2}\left(\frac{R_{\mathrm{off}}}{\sigma_{\mathrm{off}}} \right)^2 \right]\,,
\end{equation} where the parameter $\sigma_{\mathrm{off}}$ is the peak of the offset distribution.

The resulting mean surface mass profile for incorrectly centred combinations of clusters can be written then as 

\begin{equation}
\Sigma^s (R) = \int_0^\infty \mathrm{d} R_{\mathrm{off}} P(R_{\mathrm{off}}) \Sigma_{\mathrm{off}} (R | R_{\mathrm{off}})\,. 
\end{equation}

\subsubsection{Contributions from the large scale structure of the universe}

The NFW profile is expected to be a good representation of halo profiles only to a certain scale, at most $\sim 2 \Mpc/h$. To go further to outer radii, the contribution of the large scale structure must be accounted for. To do so, we can write the two halo mass contribution as

\begin{equation}
 \rho_{2h} = b(\nu) \overbrace{\Omega_m \rho_{c,0} (1+z)^3}^{\bar \rho_m (z)} \xi^L(r,z)\,,
\end{equation}where $b(\nu)$ is the linear halo bias calculated at the density peak height of the halo $\nu = \delta_c/\sigma(M)$, where $\sigma(M)$ is the linear matter variance in the Lagrangian scale of the halo, that is, with $R=\left[3M/(4\pi \bar{\rho}_m)\right]^{1/3}$.

The linear matter correlation function at redshift $z$ can  as

\begin{equation}
  \xi^L(r,z) = D(z)^2 \ \sigma_8^2 \ \xi^L_n\left[(1+z)r\right]\,,
\end{equation} where $\xi^L_n(r)$ is the linear correlation function at redshift zero, taken from a {\sc camb}-calculated \citep{CAMB} linear power spectrum, normalised to $\sigma_8=1$, and $D(z)$ is the growth function. 

Now, to calculate the projected density due to the large scale structure, we can use the bias for a halo of mass $M$ at redshift $z$ as 

\begin{equation}
 B(z,M) := b(z,M) \ \Omega_{\mathrm{M}} \ \sigma_8^2 D(z)^2\,,
\end{equation} so that we can write the projected 2-halo term as

\begin{equation}
 \Sigma_{2h}(R) = B(z,M) \Sigma_l(R)\,,
\end{equation} with

\begin{equation}
 \Sigma_l = (1+z)^2 \rho_{c,0} W((1+z)R)\,,
\end{equation} and
 
 \begin{equation}
  W(R) := \int_{-\infty}^{\infty} \de y \xi_l \left(\sqrt{y^2 + R^2}\right)\,.
 \end{equation}
 
 We have now all ingredients to our parametric model, which can use up to 5 parameters, namely $M_0, M_{200}, c_{200}, \sigma_{\mathrm{off}}, p_{\mathrm{cc}}$ - which represent, respectively, the baryonic mass of the BCG, the halo mass and concentration inside $r_{200}$, the peak of the miscentering offset radius distribution, and the fraction of systems that were miscentered in the stack.
 
\subsection{Stacking Galaxy Systems}

To address the low SNR from shape measurements, mostly due to intrinsic shape dispersion of galaxies, we use the cross-correlation lensing \citep{Johnston2007}, which consists of stacking the signal of a number of systems selected by a set of properties, such as redshift and richness. We define three redshift \emph{slices}, one at a lower redshift interval $\left(0.2<z<0.4\right)$, another in an intermediate range $\left(0.4<z<0.6\right)$, and a last one at $\left(0.6<z<0.7\right)$. Furthermore, we restrict ourselves to an interval of cluster richness $15 < \lambda < 45$, and then divide into sets of different $\Delta M_{1-2}$. We also reject systems with poorly-defined central galaxies by requiring that the probability that the CG is correctly identified (which is an output from redMaPPer) is greater than $p_{cen}>0.8$.

To compare stacks of different median magnitude gaps optimising our signal, we did not divide into "Fossil" and "non Fossil" systems, as the FS count was too low. Instead we define roughly equal-sized partitions of each redshift slice ranked by $\Delta M_{1-2}$. 

In order to test the stability of our results to different partitions, we used both terciles and quartiles cuts in each redshift slice. Additionally, we tested the higher magnitude gap tercile/quartile against a stack made from the combined rest of the systems in that redshift slice, to have a simpler picture from which we can understand our results. The resulting median values for the magnitude gap ($\Delta M_{1-2}$), richness ($\lambda$), and redshift ($z$) for each stack can be seen in Table \ref{tab:barstacks}, for the quartiles case. 

\begin{table}
\setlength{\tabcolsep}{4.5pt}
    \centering
    \bgroup
    \def\arraystretch{1.5}
    \begin{tabular}{c}

\begin{tabular}{c|c|c|c|c|c}
     &&& $\Delta M_{1-2}$ stack & \\
redshift &  & Large & Interm. I & Interm II & Small  \\ \hline
 &$N$ & $40$  & $41$  & $41$ & $42$  \\ 
$0.2<z<0.4$ &$\Delta M_{1-2}$ & $1.7^{+0.5}_{-0.1}$ & $1.1^{+0.2}_{-0.1}$ & $0.8^{+0.1}_{-0.1}$& $0.5^{+0.2}_{-0.2}$ \\ 
 &$\lambda$ & $20^{+10}_{-4}$  & $22^{+6}_{-4}$  & $22^{+11}_{-6}$& $22^{+10}_{-5}$ \\ 
 &$z$ & $0.32^{}_{}$ & $0.31^{}_{}$  & $0.33^{}_{}$& $0.34$ \\ 
\hline  \end{tabular} \\

\begin{tabular}{c|c|c|c|c|c}
&&& $\Delta M_{1-2}$ stack & \\
redshift &  & Large & Interm. I & Interm II & Small  \\ \hline
&$N$ & $82$  & $82$  & $82$& $84$  \\ 
$0.4<z<0.6$&$\Delta M_{1-2}$ & $1.7^{+0.4}_{-0.2}$ & $1.2^{+0.2}_{-0.2}$ & $0.9^{+0.2}_{-0.2}$& $0.5^{+0.2}_{-0.3}$\\ 
&$\lambda$ & $22^{+7}_{-6}$  & $20^{+12}_{-4}$  & $22^{+12}_{-5}$& $19^{+10}_{-3}$ \\ 
&$z$ & $0.50^{}_{}$ & $0.51^{}_{}$  & $0.51^{}_{}$& $0.53^{}_{}$  \\ 
\hline  \end{tabular} \\

\begin{tabular}{c|c|c|c|c|c}
&&& $\Delta M_{1-2}$  stack& \\
redshift &  & Large & Interm. I & Interm II & Small  \\ \hline
&$N$ & $111$  & $112$  & $111$ & $113$  \\ 
$0.6<z<0.7$&$\Delta M_{1-2}$ & $1.6^{+0.5}_{-0.4}$ & $1.0^{+0.3}_{-0.2}$ & $0.6^{+0.5}_{-0.3}$& $0.5^{+0.4}_{-0.5}$\\ 
&$\lambda$ & $21^{+8}_{-4}$  & $20^{+12}_{-4}$  & $20^{+11}_{-4}$& $21^{+10}_{-4}$ \\ 
&$z$ & $0.67^{}_{}$ & $0.68^{}_{}$  & $0.68^{}_{}$& $0.69^{}_{}$  \\ 
\hline  \end{tabular}

    \end{tabular}
    \egroup
    \caption{Stacks, with  the number of systems $N$, and the respective median of magnitude gaps $\Delta M_{1-2}$, richness $\lambda$, and redshift $z$ when the redshift slices are further separated into quartiles of different magnitude gaps.}
    \label{tab:barstacks}
\end{table}

Having defined the stacks of similar systems, we combine the lensing signal of each stack by first defining radial annuli, and calculating the estimator for the lensing signal for each ring as:

\begin{equation}\label{eq:deltasigma}
    \widehat{\Delta \Sigma} (R) = \frac{\sum_{d,s} w^{d,s} \sigmacrit^{d,s}   {\gamma_t}^{s}}{\sum_{d,s}w^{d,s}} 
\end{equation} where $\gamma_t$ is the tangential ellipticity of each background source galaxy, $\sigmacrit^{d,s}$ is the lensing critical density for each pair system ($d$) galaxy ($s$) and $w^{d,s}=w^{s}\left(\sigmacrit^{d,s}\right)^{-2}$ is the lensing weight, which quantifies the measurement error in each source galaxy.   

Errors in the lensing signal estimator were calculated by bootstrapping each stack $N=200$ times, from which we also calculate the full covariance matrix between each pair of rings. 

\begin{multline}
    \mathcal{C}_{i,j} = \left[\frac{N}{N-1} \right]^2 \frac{1}{N} \sum^N_k \left[\Delta \Sigma_k(R_i)-\overline{\Delta \Sigma_k}(R_i) \right] \times \\ \left[\Delta \Sigma_k(R_j)-\overline{\Delta \Sigma_k}(R_j) \right] 
\end{multline}

We have tested several values ($N=50,150,200,300$) for the bootstrap number and arrived at roughly similar error bars in all of them, without any clear trend of diminishing errors with higher $N$. Also, bootstrap errors were compared to simple standard deviation of the shear signal in each bin, resulting in the standard deviation being a less conservative (smaller error bars) approach.

\section{Analysis}

We have calculated the lensing signal $\Delta \Sigma(R)$ for $6$ logarithmically spaced rings, spanning from about $100 \mathrm{kpc}$ to $10 \mathrm{Mpc}$, and using the geometric centre of each radial bin as values for $R$. Using the {\sc emcee} code of \citet{ForemanMackey2013}, we find posterior distributions for the parameters in our model by performing an MCMC fit of a multivariate Gaussian likelihood given by

\begin{multline}
\ln{\mathcal{L}} = -\frac{1}{2}\left[ \left(\widehat{\Delta \Sigma} -\Delta \Sigma(R|M_{200},c_{200},p_{\mathrm{cc}}) \right)^T \times C^{-1} \times \right. \\
\left. \left(\widehat{\Delta \Sigma}- \Delta \Sigma(R|M_{200},c_{200},p_{\mathrm{cc}})\right)  \right] \,,
\end{multline} where $C^{-1}=\mathcal{H}\mathcal{C}^{-1} $ is the inverse of the covariance matrix corrected by the method of \citet{Hartlap}. We use flat priors for concentration and mass, and a Gaussian prior on $p_{\mathrm{cc}}$ as shown in table \ref{tab:priors}. The prior on the correctly centred fraction comes from the redMaPPer algorithm, which gives probabilities for the correct identification of the CG. We discarded the baryonic component of the central galaxy and used a fiducial value for the peak of the centre offset distribution after tests had shown that fitting the full model did not affect our results other than resulting in less precision in $\widehat{\Delta \Sigma}$ errors.

\begin{figure}
 \includegraphics[width=\columnwidth]{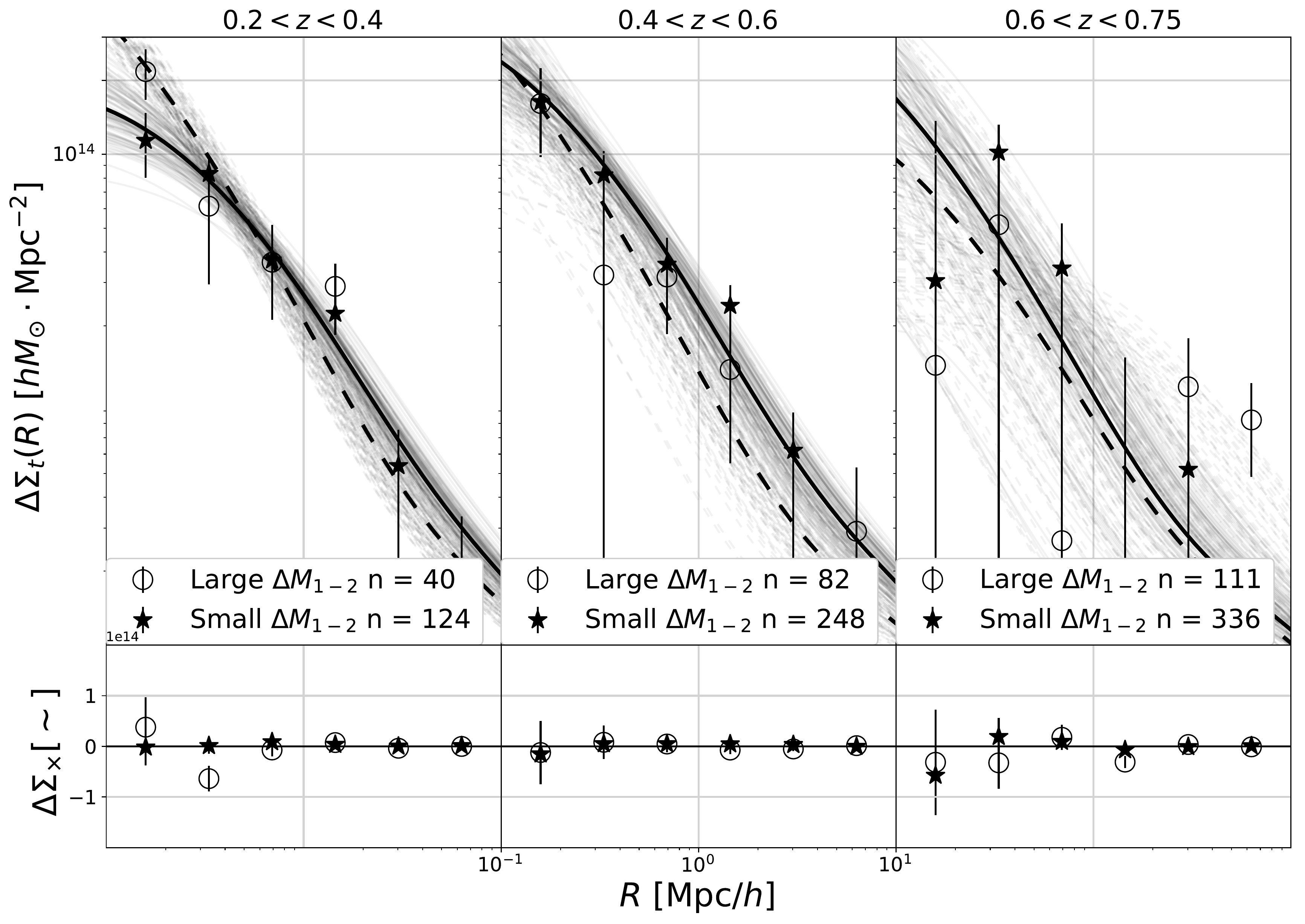}
 \caption{Lensing signal data points combined with lines that represent model fitting. The thinner, diffuse, lines represent a random selected sample of chain links of the posterior probability chain obtained by the MCMC - which serves as a picture of uncertainty, while the thicker line represents the median-derived best values for the model. The cross-ellipticity shown below is consistent with no systematic effects.}
 \label{fig:results}
\end{figure}

The results of the $\Delta \Sigma(R)$ estimation and the fitting process are shown together in figure \ref{fig:results}, where we plot the result for a stack of the quartile of the largest $\Delta M_{1-2}$ systems against a stack of the others in the three redshift slices. The plot shows both the tangential and cross ellipticity components, the latter being a tool to diagnose any systematic effect. 

\begin{table}
    \centering
    \bgroup
    \def\arraystretch{1.5}    
    \begin{tabular}{c|c}
    Parameter     &  Prior \\ \hline
    $\log\left(M_{200}\right)$ &    $[12,15]$    \\
    $c_{200}$     &  $[0.1,20]$       \\
    $p_{\mathrm{cc}}$      &  $\mathcal{N}(\overline{p_{\mathrm{cen}}},\sigma_{p_{\mathrm{cen}}})$      \\
    $\sigma_{\mathrm{off}}$&  $0.41 \  \Mpc/h$       \\
    $M_0$         &  $0$   \\ \hline
    
    \end{tabular}
    \egroup
    \caption{Priors used with the MCMC fitting process. The $\log\left(M_{200}\right)$ and $c_{200}$ are flat priors. The $p_{\mathrm{cc}}$ prior is determined by a Gaussian with centre and width determined by the redMaPPer given $p_{\mathrm{cen}}$ of the highest probability central galaxy for each system. We have also fixed the miscentering offset parameter to a fiducial value and the BCG baryonic mass to zero, as the data have not enough SNR to constrain a full parameter model.}
    \label{tab:priors}
\end{table}

We can see in figure \ref{fig:results} that the slope of the mass profile is steeper in the centre for the stacks with larger magnitude gaps, especially in the lower redshift slice, and that the high redshift slice is poorly constrained, which can be attributed to shortcomings in the cluster finder and the observational depth.

\subsection{Posterior Distributions for Masses and Concentrations}

\begin{figure}
 \includegraphics[width=\columnwidth]{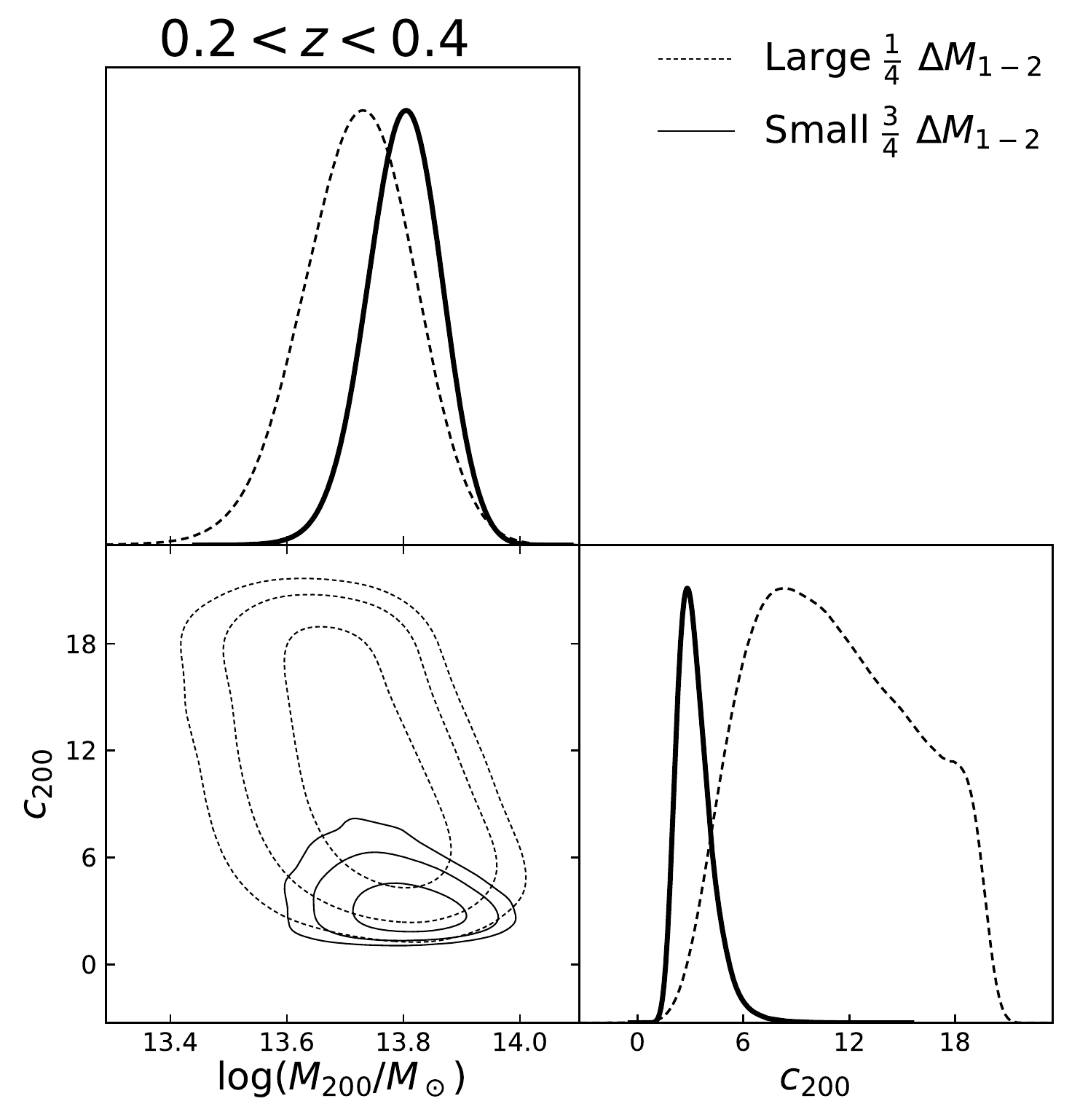}
 \caption{MCMC-computed $\mathrm{2D}$ posterior distributions for lower ($0.2<z<0.4$) redshift stacks with peak-normalised $\mathrm{1D}$ posteriors of mass and concentration, taken from the quartile partition, comparing the higher magnitude gap quartile with a stack with the other three. The contour lines progressively define $68\%$, $95\%$ and $99\%$ confidence level regions.}
 \label{fig:trianglethirds0}
\end{figure}

\begin{figure}
 \includegraphics[width=\columnwidth]{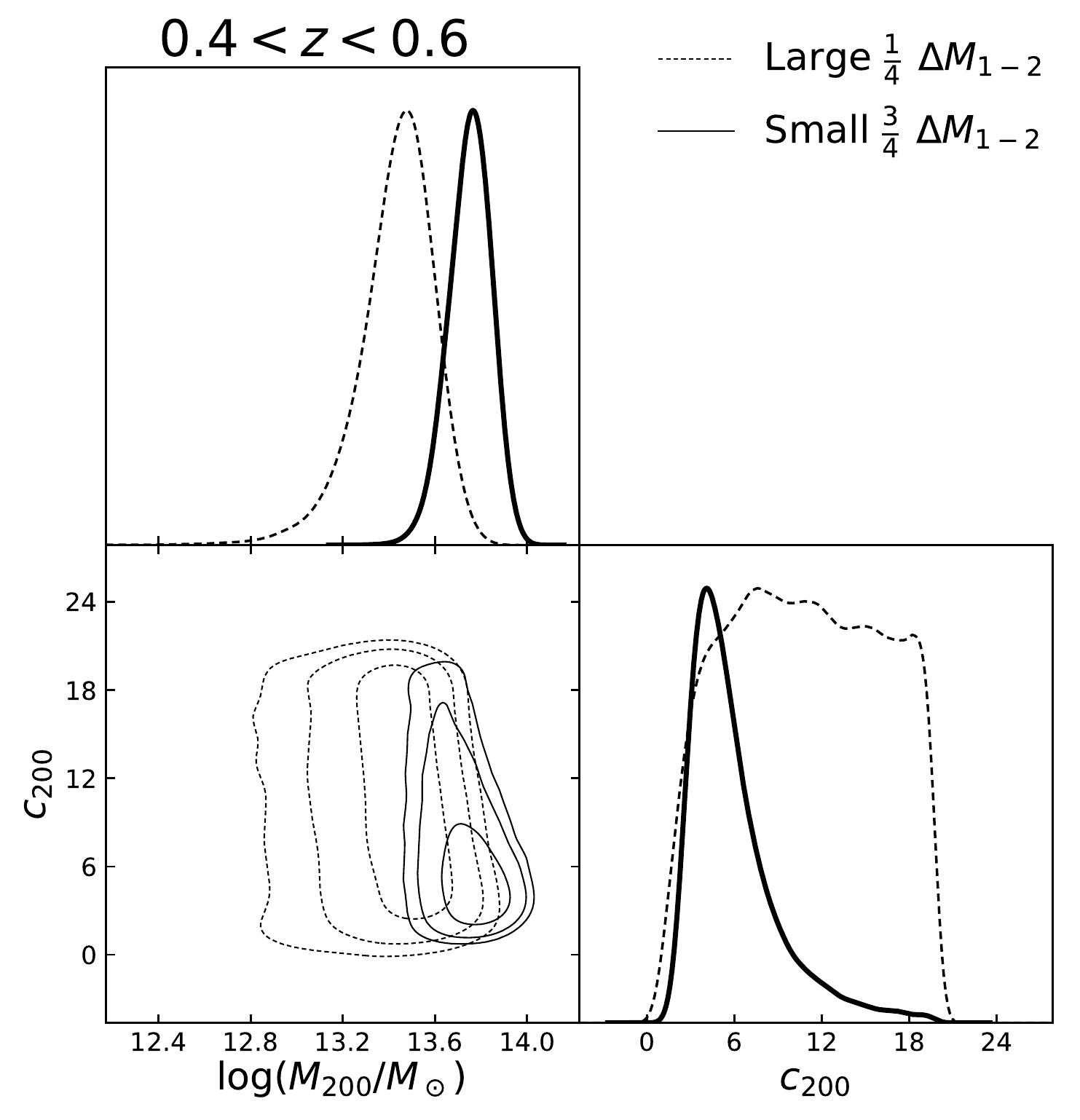}
 \caption{MCMC-computed $\mathrm{2D}$ posterior distributions for lower ($0.4<z<0.6$) redshift stacks with peak-normalised $\mathrm{1D}$ posteriors of mass and concentration, taken from the quartile partition, comparing the higher magnitude gap quartile with a stack with the other three. The contour lines progressively define $68\%$, $95\%$ and $99\%$ confidence level regions. The concentration of the large magnitude gap stack is not well resolved, a situation that is prevalent among the higher redshift stacks, which prevented further analysis.}
 \label{fig:trianglethirds1}
\end{figure}

We present the 2D posterior distributions marginalised over $p_{\mathrm{cc}}$ in figures \ref{fig:trianglethirds0} \& \ref{fig:trianglethirds1}. In the intermediate redshift slice ($0.4<z<0.6$) we have been unable to constrain the concentration for the quartiles case, with the posterior spreading flat as the prior. In the third (highest) redshift slice, both masses and concentrations are poorly constrained by the data, with the posterior spread through the prior, and we restrain from further analysis.  

The mass posteriors of each stack in the two lower redshift slices have medians of the NFW $M_{200}$ that are in agreement with expectations for the richness $\lambda$ of each stack \citep{redMaPPer}.

\subsection{Correcting for the mass-concentration relation}
\begin{center}
\begin{table*}
\setlength{\tabcolsep}{1pt}
    \centering
    \bgroup
    \def\arraystretch{1.5}
    \begin{tabular}{c|c}
Terciles & Quartiles \\

\begin{tabular}{c|c}

$0.2<z<0.4$ & $0.4<z<0.6$ \\
\setlength{\tabcolsep}{2.0pt}
\begin{tabular}{c|c|c|c}

$\Delta M_{1-2}$ & Large & Interm. & Small  \\ \hline  \\[-1em] 
$M_{200}$ & $5.1^{+1.4}_{-1.2}$ & $5.8^{+1.5}_{-1.3}$ & $4.5^{+1.2}_{-1.0}$   \\ 
$c_{200}$ & $6.3^{+4.6}_{-2.3}$  & $3.1^{+1.7}_{-1.1}$  & $6.2^{+6.4}_{-3.0}$    \\ 
$c_{200}^{\mathrm{corr}}$ &$6.3^{+4.6}_{-2.3}$  & $3.2^{+1.7}_{-1.1}$  & $6.1^{+6.1}_{-3.0}$   \\ 
$p_{\mathrm{cc}}$ & $0.96$  & $0.95$  & $0.92$    \\ 
\hline  \end{tabular} &

\begin{tabular}{c|c|c}
Large & Interm. & Small  \\ \hline  \\[-1em]
 $4.5^{+1.2}_{-1.0}$ & $2.7^{+1.0}_{-0.8}$ & $4.4^{+1.5}_{-1.3}$ \\ 
$11.4^{+5.7}_{-5.6}$  & $5.3^{+6.5}_{-2.9}$  & $4.6^{+3.8}_{-1.8}$   \\ 
$11.4^{+5.7}_{-5.6}$ & $5.6^{+6.6}_{-3.0}$  & $5.1^{+4.1}_{-2.0}$   \\ 
$0.95$  & $0.94$  & $0.90$\\ 
\hline  \end{tabular}

\end{tabular} &

\begin{tabular}{c|c}

$0.2<z<0.4$ & $0.4<z<0.6$ \\

\begin{tabular}{c|c|c|c}
 Large & Interm. I & Interm II & Small  \\ \hline  \\[-1em] 
$5.1^{+1.2}_{-1.0}$ & $5.3^{+1.5}_{-1.4}$ & $7.7^{+2.0}_{-1.8}$ & $3.9^{+1.3}_{-1.1}$ \\ 
 $11.0^{+5.5}_{-4.4}$  & $4.5^{+4.0}_{-1.9}$  & $1.7^{+0.6}_{-0.5}$&  $8.1^{+6.9}_{-4.3}$  \\ 
$11.0^{+5.5}_{-4.4}$  & $4.5^{+3.8}_{-2.0}$  & $1.8^{+0.7}_{-0.5}$& $7.7^{+6.5}_{-4.2}$  \\ 
$0.96$  & $0.95$  & $0.95$ &    $0.92$     \\ 
\hline  \end{tabular} &

\begin{tabular}{c|c|c|c} 
Large & Interm. I & Interm II & Small  \\ \hline  \\[-1em]
$3.1^{+1.1}_{-1.0}$ & $3.7^{+1.5}_{-1.3}$ & $3.3^{+1.6}_{-1.4}$ & $10.0^{+2.5}_{-2.2}$\\
$10.0^{+6.4}_{-5.5}$  & $7.2^{+7.4}_{-4.3}$  & $8.7^{+7.1}_{-5.3}$ & $4.8^{+3.1}_{-1.7}$   \\ 
$10.0^{+6.4}_{-5.5}$ & $7.4^{+7.4}_{-4.4}$  & $8.7^{+7.0}_{-5.3}$ & $5.3^{+3.4}_{-1.9}$   \\ 
$0.96$  & $0.95$  & $0.93$ &    $0.90$  \\ 
\hline  \end{tabular} 
\end{tabular}
\end{tabular}
\egroup
\caption{Posterior medians and $68\%$CL intervals for the masses, concentrations, corrected concentrations and miscentered fraction of systems, each calculated by marginalising other parameters. Masses are given in units of $10^{13} ~\mathrm{M}_\odot$, and the $68$CL intervals around the median $p_{\mathrm{cc}}$ values are of about $\sim \pm 0.05$.}
    \label{tab:results2}
\end{table*} 
\end{center}

In the hierarchical scenario of the formation of the large scale structure of the universe, more massive halos assemble latter and thus, are expected to be less concentrated \citep{NFW97}. This gives rise to a mass-concentration relation where more massive objects are generally less concentrated. Therefore, to compare stacks of different masses properly, we must first correct the concentrations to compare them on an ``equal mass'' reference. 

The mass-concentration relation of dark matter halos has been studied by several authors over many studies. We have used the relation given by \citet{Duffy08} (Eq. \ref{duffy}) as a scaling relation to offset the concentrations to of the lower $\Delta M_{1-2}$ to the higher one. 

\begin{equation}\label{duffy}
    c_{200}^{i,\mathrm{corr}} = c_{200}^i\left( \frac{M_{200}^j}{M_{200}^i}\right)^\beta ~ \mathrm{with} ~ \beta = -0.091 \,.
\end{equation}

\begin{figure}
 \includegraphics[width=\columnwidth]{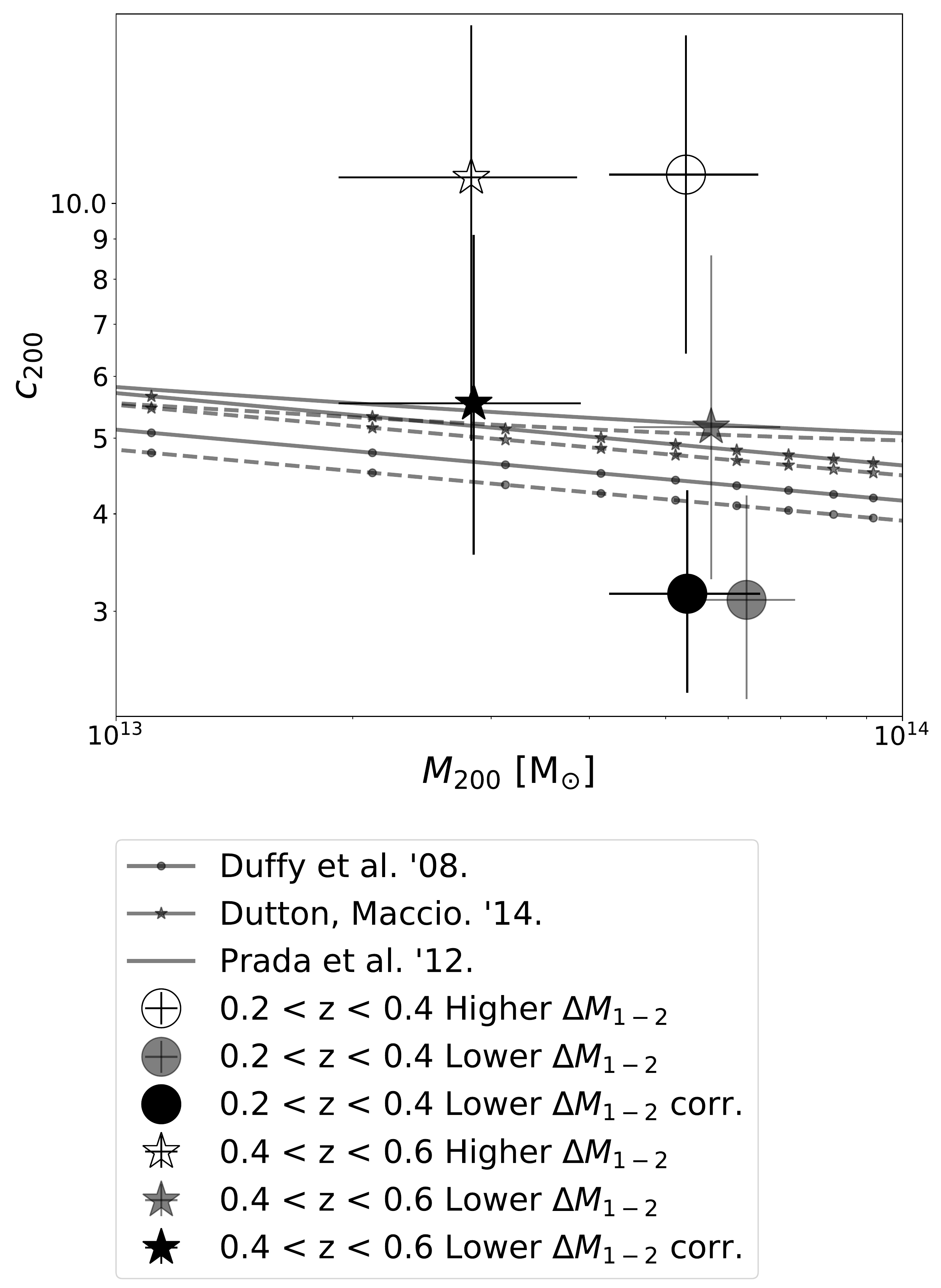}
 \caption{Adjusting concentrations to compare in an "equal-mass" footing, we slide them to the left using the relation of \citet{Duffy08}. The solid markers are corrected  mass-concentrations of the MCMC-calculated translucent translucent ones. The lines represent several $c-M$ relations for $z=0$ and the dashed for $z=0.5$. This graph corresponds to the comparison between the largest $\Delta M_{1-2}$ stack against the stack of the rest three quartiles of systems in each redshift slice.}
 \label{fig:duffycm4}
\end{figure}

In order to apply this correction taking into account the full probability distribution in mass and concentration, we calculate $c_{200}^{corr}$ by choosing random pairs of the MCMC chain links from the reference stack (the largest $\Delta M_{1,2}$ one) and the one to be corrected, and apply the scaling relation from the Duffy relation. We have then a probability distribution for the corrected concentrations. The resulting shift in concentration can be seen in figure \ref{fig:duffycm4}, where we show the correction in mass and concentration for the Large $\Delta M_{1-2}$ quartile against the stack of the other 3 quartiles, in both the low and medium redshift slices.

To summarise the results of the MCMC fitting we present in table \ref{tab:results2} the median and $68\%$CL intervals for the halo mass ($M_{200}$), concentration $c_{200}$, corrected concentration $c_{200}^{\mathrm{corr}}$, and $p_{\mathrm{cc}}$ for each tercile and quartile from the two lower redshift slices. 

\section{Results}

\begin{figure*}
 \includegraphics[width=2\columnwidth]{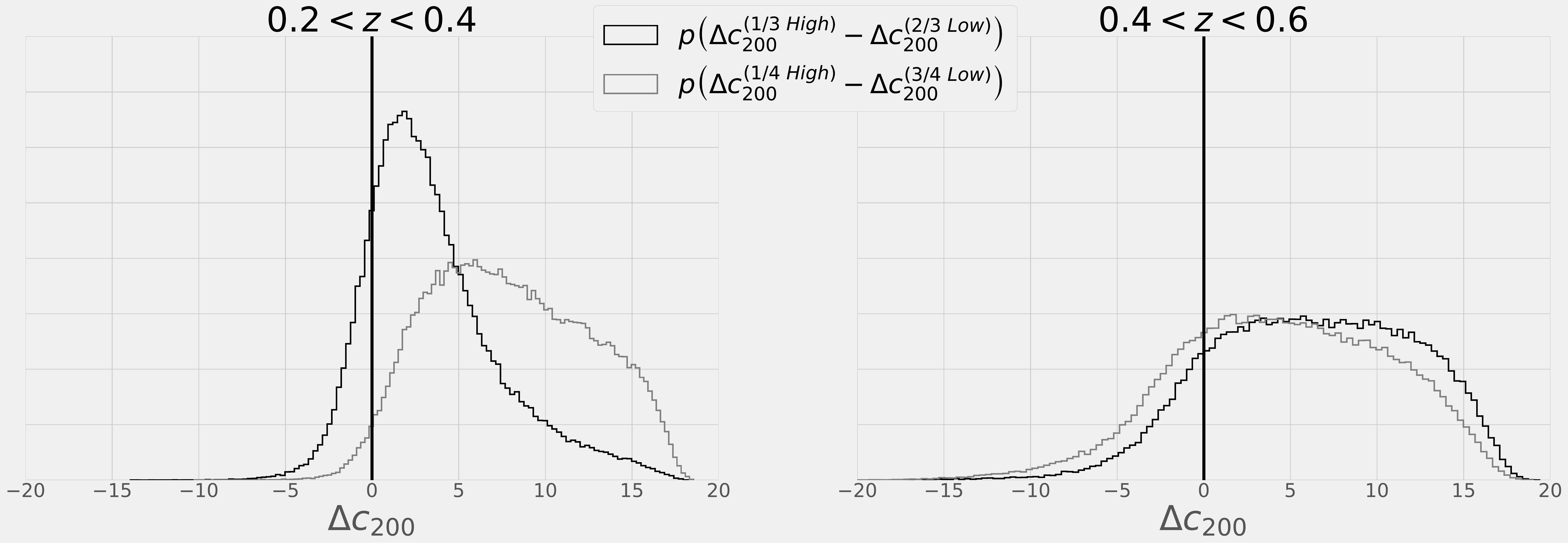}
 \caption{Probability distribution for the difference in concentration ($\Delta c_{200}$) between larger and smaller $\Delta M_{1-2}$ stacks, when each redshift interval is divided into terciles (black) and quartiles (grey) of different magnitude gaps, using the 1 versus the rest comparison. The probabilities that the larger magnitude gap stack is more concentrated than the smaller is then given by the portion of the distribution left to the $\Delta c_{200}=0$ line.}
 \label{fig:diff_fourths}
\end{figure*}

We have quantified the differences in corrected concentrations between different stacks by building, for each pair in a redshift slice, a chain of differences between randomly selected chain links of each corrected concentration chain. This resulting chain samples the probability distribution for the differences $\Delta c_{200}$ and the results can be seen in figure \ref{fig:diff_fourths} again for the case of 1 quartile of large $\Delta M_{1-2}$ against the other three, in both low and medium redshift slices.

By integrating the positive side of these distributions (which can be simply done by counting the fraction of chain links with $\Delta c_{200}>0$) we calculate probabilities that one be greater than the other, given the previous concentration results. We display these results in table \ref{tab:delta_c}.

As shown in tables \ref{tab:delta_d} \& \ref{tab:delta_c}, for the lower redshift slice ($0.2<z<0.4$) we find that the larger magnitude gap stack has a high probability of being more concentrated than the stack of other systems both when divided in terciles ($P=0.84$), and quartiles ($P=0.98$).

In the intermediate redshift slice ($0.4<z<0.6$) we again see the same tendencies, where the first terciles (quartiles) have probabilities of $P=0.85$($P=0.76$) of being more concentrated than the other systems stacked together. As we have seen in figure \ref{fig:trianglethirds1}, however, we have to be cautious to consider this result, as the concentration for the higher magnitude gap quartile was not well defined, with its probability distribution being mostly flat as the prior. In the terciles, however, the posterior was much better defined.   

\begin{table}
    \centering
    \begin{tabular}{clcc}
        redshift & Terciles & $\Delta c_{200}$ & $P\left(\Delta c_{200} >0\right)$\\
         \hline
         \hline
                 &``Large-Intermediate'' & $3.1^{+4.5}_{-2.8}$ &  $0.86$\\ 
    $0.2<z<0.4$    &``Large-Small''        & $0.3^{+4.9}_{-6.0}$ &  $0.52$\\
        &``Intermediate-Small'' & $-2.8^{+3.4}_{-6.1}$ &  $0.20$\\
    \hline 
    $0.2<z<0.4$    &``Large-Others''        & $2.8^{+4.3}_{-2.8}$ &  $0.84$\\
             \hline  
      $0.4<z<0.6$  &``Large-Others'' & $6.3^{+6.2}_{-6.1}$ &  $0.85$\\
    \end{tabular}
\caption{Median values with $68\%$CL intervals for the difference in concentration between tercile stacks by magnitude gap in the two lower redshift slices.}    
\label{tab:delta_d}

    \begin{tabular}{clcc}
       redshift & Quartiles & $\Delta c_{200}$ & $P\left(\Delta c_{200} >0\right)$\\
         \hline
         \hline
           &``Large-Interm. I'' & $5.9^{+5.9}_{-5.4}$ &  $0.86$\\ 
  
        &``Large-Interm. II''& $9.1^{+5.5}_{-4.5}$ &  $>0.99$\\
      $0.2<z<0.4$  &`Large-Small''' & $2.8^{+6.5}_{-6.8}$ &  $0.67$\\
        &``Interm. I-Interm. II'' & $2.6^{+3.9}_{-2.1}$ &  $0.92$\\ 
    &``Interm. I-Small''        & $-2.9^{+5.1}_{-6.7}$ &  $0.29$\\
        &``Interm. II -Small'' & $-5.9^{+4.2}_{-6.5}$ &  $0.04$\\
         \hline  
      $0.2<z<0.4$  &``Large-Others'' & $7.6^{+5.5}_{-4.5}$ &  $0.98$\\
               \hline  
      $0.4<z<0.6$  &``Large-Others'' & $4.5^{+6.6}_{-6.1}$ &  $0.76$\\
 
    \end{tabular}

    \caption{Median values with $68\%$CL intervals for the difference in concentration between quartile stacks by magnitude gap in the two lower redshift slices.}
    \label{tab:delta_c}
\end{table}

\section{Discussion}

In general, our data indicate that populations of galaxy systems with larger magnitude gaps have, on average, more concentrated mass profiles.  In our analysis we did not attempt to segregate Fossil Systems into a particular stack, to optimise our lensing SNR. However, we still managed to see a trend of larger magnitude gap systems having more concentrated halos. The difference in concentrations between stacks was most significant when comparing the higher magnitude gap stack against the rest of the systems. Some of this trend still appears in the intermediate redshift slice $0.4<z<0.6$ though, unlike in the low redshift slice, the results were not statistically significant to the $95\%$CL. 

We have found some pairs of stacks in which this trend is reversed. This, however, only happens in smaller magnitude gap pairs, where the difference in median $\Delta M_{1-2}$ of each of the stacks is smaller. We expect then that the systems are less different in nature and these particular results can be due to random scatter.

Apart from statistical fluctuations, it is possible that the smallest $\Delta M_{1-2}$ stack of each redshift slice have more systems with misidentified CGs, and ongoing mergers. This not only complicates the choice of centre, but also can bias parametric mass estimates \citep{Hoekstra2002,Hoekstra13}.

Possible problems in our analysis can arise due to biased sampling, since cluster finding algorithms can have preconceptions on what a galaxy cluster/group \emph{should} look like. The redMaPPer algorithm, for instance, assumes a Schechter-like luminosity function, which can undercount more extreme $\Delta M_{1-2}$ systems. In fact, we find a very low fraction of systems as Fossils($\sim 5\%$), in comparison with the literature ($\sim 10 - 20\%$) which may be a failure of the algorithm to identify systems with large magnitude gaps.

Nevertheless, our results push further the argument that the magnitude gap is an indicator of early forming systems, as there is theoretical basis for the early formation of more concentrated systems \citep{NFW95}. This shows that, despite the evolution of magnitude gaps and contamination by younger systems, that is expected from simulations, we may still see some information on mass. Further confirmation will come when more statistically complete populations of galaxy systems become available in future surveys.

\section{Summary and Conclusions}

We have performed the first weak lensing mass distribution analysis of galaxy systems ranked by the magnitude gap of the central galaxy (CG) and the brightest satellite galaxy (BSG). Using cross-correlation lensing of stacked systems, divided by redshift and rank of $\Delta M_{1-2}$ we calculated thee shear signal in radial bins and, to that data, applied a MCMC procedure to calculate posterior distributions for the mass and concentrations of these stacks. After correcting the concentration for mass differences by a $c-M$ relation, we calculated the probability distribution of differences in concentrations between the stacks in each redshift slice, dividing them both in terciles and in quartiles. Finally, we integrate the positive side of these distributions to calculate a probability of the larger $\Delta M_{1-2}$ stack having a more concentrated halo according to our data.

We have found significant evidence in our data indicating that populations of systems with larger magnitude gaps are more likely to have more concentrated mass distributions. Assuming that our model describes the mass distributions sufficiently well, we find that, in the lower redshift slice ($0.2<z<0.4$), our larger magnitude gap quartile stack is more concentrated than the rest at more than $95\%$CL. We also find the same trend in the intermediate redshift slice ($0.4<z<0.6$) although with less confidence.

These results agree with and strengthen the claim of larger magnitude gaps being correlated to more concentrated halos \citep{Khosroshahi2007}. This, together with investigations through simulations by \citet{Deason2013,Gozaliasl14} further the claim that a substantive fraction of systems with larger magnitude gaps have been formed early in the history of the Universe. As these simulations are all based in the $\Lambda$CDM scenario of cosmology, these results are compatible with the same assumptions. Recent work \citep{Gozaliasl14} with simulations further indicate that magnitude gaps can be used together with other tracers to efficiently discriminate virialised populations of galaxy systems. These more relaxed systems offer, in turn, more insight into Cosmology and the formation of the large scale structure of the universe \citep{Mantz2015}.  

Finally, large upcoming optical surveys will present more complete populations of galaxy systems at low redshifts in great areas of the sky such as  JPAS \citep{JPAS}, Euclid \citep{euclid}, and others. With more precise photometric redshifts as in the case of JPAS, weak lensing will be at an advantage point to look for more evidence of the dynamical nature of large magnitude gap systems. 

\section{Acknowledgements}

This work is based on observations obtained with MegaPrime/MegaCam,a joint project of CFHT and CEA/DAPNIA, at the Canada-France-Hawaii Telescope (CFHT),which  is  operated  by  the  National  Research  Council (NRC) of Canada, the Institut National des Science de l’Univers of the Centre National de la Recherche Scientifique (CNRS) of France, and the University of Hawaii. The Brazilian partnership on CFHT is managed by the Laboratorio Nacional de Astrofısica  (LNA). This work  made  use  of  the  CHE  cluster,  managed  and  funded by ICRA/CBPF/MCTI, with financial support from FINEP and FAPERJ. We thank the support of the Laboratorio Interinstitucional de e-Astronomia (LIneA). We thank the CFHTLenS team for their pipeline development and verification upon which much of this surveys pipeline was built.

M.~Makler is partially supported by CNPq and FAPERJ. Fora Temer. The authors would also like to acknowledge support from the Brazilian agencies CNPq (E. Cypriano, A. Vitorelli), FAPESP (2014/13723-3) (E. Cypriano) and CAPES (A. Vitorelli). 

The authors would like to thank Renato Dupke, Claudia Mendes de Oliveira, and Aldée Charbonnier for useful comments and discussions.

\bibliographystyle{mnras}
\bibliography{clusters,cosmo,lensing,fossil,surveys}

\appendix

\bsp	
\label{lastpage}
\end{document}